\documentstyle[galley, rotate]{mn}
\onecolumn
\input{epsf}

\def\inc{{\int_0^{\chi_s}}}

\def\kmin1{{\int_0^{\chi_s} \omega(\chi) d \chi }}
\def\corr{\left [ \int  {d^2 {\bf l} \over (2
\pi)^2 )} {\rm  P}( { \bf l \over r(\chi) })  W^2(l\theta_0) \exp [ i l
\theta_{12}] \right ]}
\def\corri{\left [ \int  {d^2 {\bf l} \over (2
\pi)^2 )} {\rm  P}( { \bf l \over r(\chi) })  W^2(l\theta_0) \exp [ i l
\theta_{23}] \right ]}
\def\corrj{\left [ \int  {d^2 {\bf l} \over (2
\pi)^2 )} {\rm  P}( { \bf l \over r(\chi) })  W^2(l\theta_0) \exp [ i l
\theta_{34}] \right ]}
\def\var1{ \Big [  \int  {d^2 {\bf l} \over (2
\pi)^2 )} {\rm P} \left ( { \bf l \over r(\chi) } \right )
W^2(l\theta_0) \Big ] }
\def\var {\kappa_{\theta_0}}
\def\corr{\kappa_{\theta_{12}}}
\def\corri{\kappa_{\theta_{23}}}
\def\corrj{\kappa_{\theta_{34}}}
\def\kmin{{\kappa_m }}

\def\one{\langle \kappa_s^2 \rangle}
\def\two{\langle \kappa_s(\gamma_1) \kappa_s(\gamma_2) \rangle}

\title[Weak Lensing]
{Cosmic Statistics through Weak Lenses}

\author[D. Munshi and  P. Coles]{Dipak Munshi$^{1}$\thanks{e-mail: mdipak@mpa-garching.mpg.de}
and Peter Coles$^2$\thanks{e-mail: Peter.Coles@nottingham.ac.uk} \\
 $^1$Max-Planck-Institut fur Astrophysik,
Karl-Schwarzschild-Str.1, D-85740, Garching, Germany\\
$^2$School of Physics \& Astronomy, University
of Nottingham, University Park, Nottingham, NG7 2RD, United
Kingdom\\
}

\pagerange{000,000}

\begin{document}
\maketitle

\begin{abstract}
Many recent studies have demonstrated that scaling arguments, such
as the so-called hierarchical {\em ansatz}, are extremely useful
in understanding the statistical properties of weak gravitational
lensing. This is especially true on small angular scales (i.e. at
high resolution), where the usual perturbative calculations of
matter clustering no longer apply. We build on these studies in
order to develop a complete picture of weak lensing at small
smoothing angles.  In particular, we study the full probability
distribution function, bias and other multipoint statistics for
the ``hot spots'' of the convergence field induced by weak
lensing, and relate these  to the statistics of overdense regions
in underlying mass distribution.  It is already known that weak
lensing can constrain the background geometry of the Universe, but
we further show that it can also provide valuable information
about the statistics of collapsed objects and the physics of
collisionless clustering. Our results are particularly important
for future observations which will, at least initially, focus on
small smoothing angles.
\end{abstract}


\section{Introduction}
The images of high redshift galaxies can be distorted by the
gravitational lensing of light passing through intervening density
fluctuations. Even the relatively weak lensing produced by
large-scale galaxy clustering can provide valuable information
about the distribution of mass on cosmological scales, as recent
feasibility studies have demonstrated (Bacon, Refregier \& Ellis
2000; van Waerbeke et al. 2000). Gravitational lensing studies
offer a particular advantage over traditional studies of
large-scale structure that rely on galaxies as tracers of mass.
The analysis of galaxy catalogues can only provide us with
information as to how galaxies are clustered, and galaxies need
not be accurate tracers of the mass distribution, owing to the
intervention of bias. To infer statistical properties of the
underlying mass distribution from galaxy catalogues one needs to
understand fully the relationship between mass and light, which is
far from trivial to unravel. Weak lensing is produced by the mass
distribution, and can therefore be used to probe the underlying
mass distribution directly without having to allow for bias
(Mellier 1999; Bernardeau 1999; Bartelmann \& Schneider 1999).

After original suggestions by Gunn (1967), pioneering
contributions on weak lensing were made by Blandford et al.
(1991), Miralda-Escud\'{e} (1991), and Kaiser (1992). More recent
developments basically follow two strands. Some authors have
focussed on the case where large smoothing angles are used. In
this regime highly non-linear effects are washed out, allowing
simpler linear and quasi-linear analyses to be performed
(Villumsen 1996; Stebbins 1996; Bernardeau et al. 1997; Kaiser
1998). A perturbative analysis can  not be used to study lensing
at small angular scales because the perturbation series begins to
diverge as this limit is approached. The other approach that has
been taken has involved the development of numerical experiments,
usually using ray-tracing methods through N-body simulations
(Schneider \& Weiss 1988; Jarosszn'ski et al. 1990; Lee \&
Paczyn'ski 1990; Jarosszn'ski 1991; Babul \& Lee 1991; Bartelmann
\& Schneider 1991; Blandford et al. 1991). Building on the earlier
work of Wambsganns et al. (1995, 1997, 1998) the most detailed
numerical study of lensing so far was performed by Wambsganns, Cen
\& Ostriker (1998). Other recent studies involving ray-tracing
experiments have been conducted by Premadi, Martel \& Matzner
(1998), van Waerbeke, Bernardeau \& Mellier (1998), Bartelmann et
al. (1998) and Couchman, Barber \& Thomas (1998).

A complete statistical analysis of weak lensing on small angular
scales is not available at present, cheifly because no
corresponding analysis exists for the underlying density field.
There are, however, several non-linear {\em ansatze} available.
These predict a tree hierarchy for the matter correlation
functions and are successful in some respects at modelling the
results from numerical simulations  of gravitational clustering
(Davis \& Peebles 1977; Peebles 1980; Fry 1984; Fry \& Peebles
1978; Szapudi \& Szalay 1993, 1997; Scoccimarro \& Frieman 1998;
Scoccimarro et al. 1998). The different {\em ansatze} involved in
these studies all involve a tree hierarchy, but disagree with each
other in the way they assign weights to trees of same order but
different topology (Balian \& Schaeffer 1989; Bernardeau \&
Schaeffer 1992; Szapudi \& Szalay 1993; Boschan, Szapudi \& Szalay
1994). The overall scaling in such models, determined by the time
evolution of the two-point correlation function, is left
arbitrary. However recent studies by several authors (Hamilton et
al 1991, Nityananda \& Padmanabhan 1994, Jain, Mo \& White 1995;
Padmanabhan et al. 1996; Peacock \& Dodds 1996) have furnished an
accurate fitting formula for the evolution of the two-point
correlation function which can be used in combination with the
hierarchical {\em ansatze} to predict all clustering properties of
the dark matter distribution in the universe.

The statistical treatments of lensing that have been carried out
so far have mainly concentrated on the properties of low-order
cumulants (van Waerbeke, Bernardeau \& Mellier 1998; Schneider et
al. 1998; Hui 1999; Munshi \& Coles 2000; Munshi \& Jain 1999a),
cumulant correlators (Munshi \& Jain 1999b) and errors associated
with their measurement from observational data (Reblinsky et al.
1999; Schneider et al. 1998). However it is well known that higher
and higher order moments are more and more sensitive to the tail
of the distribution from which they are derived represent and are
also more sensitive to errors due to finite catalogue size
(Colombi et al. 1995; Colombi et al. 1996; Szapudi \& Colombi
1996; Hui \& Gaztanaga 1998). On the other hand, numerical
simulations involving ray tracing techniques have demonstrated
that the distribution of lensed fluctuations can be of
considerable assistance in the estimation of cosmological
parameters from observational data (Jain \& Seljak 1997; Jain,
Seljak \& White 1999; Jain \& Van Waerbeke 1999). The probability
distribution function associated with density field is not
sensitive to cosmological parameters in the way its weakly-lensed
counterpart is.

Munshi \& Jain (1999a,b) and Munshi (2000) have extended these studies
to show that the hierarchical {\em ansatz} can actually be used to make
concrete analytical predictions for all statistical properties of
the convergence that results from weak lensing. Valageas (1999a)
has also used the hierarchical {\em ansatz} to compute the error
involved in estimating the cosmological parameters $\Omega_0$ and
$\Lambda_0$ from supernova observations; see alo Wang (1999). In
Valageas (1999b) the effect of smoothing was incorporated successfully
to compute the PDF of the convergence field.

In this paper we extend these analyses still further to develop a
complete picture of statistical properties of the sky-projected
density field  obtained by weak-lensing surveys. In particular, we
obtain an analytic prediction for probability distribution
function (PDF), bias and higher-order moments for regions where
the projected density is particularly high. We also show how these
the statistics of these `hotspots' are related to similar
quantities in the underlying mass distribution. These results can
 be used to understand the effect of changing source redshift,
smoothing angle and  cosmological parameters such as $\Omega_0$
and $\Lambda_0$. This will allow a more detailed exploration of
cosmological parameter space than is possible with ray-tracing
experiments. Several of the analytic results we discuss here have
already been tested against high resolution numerical simulations
and found to be in very good agreement. Finally, we will also show
that our present understanding of gravitational clustering is
sufficient  to make firm predictions that can be tested using
future observations.

Throughout this analysis we will be neglecting noise due to source
ellipticity and will also be neglecting source clustering. A
complete error analysis of different statistics with these
contributions taken into account will be presented elsewhere
(Munshi \& Coles 2000b).

The paper is organized as follows. In Section 2 we present a
detailed analysis of the PDF for smoothed convergence field
$\kappa_s$. In Section 2 we present an analysis of the bias
associated with peaks in smoothed convergence field $\kappa_s$.
Sections 4-6 are devoted to the study of third-, fourth-, and
fifth-order statistics (respectively) of the smoothed convergence
field. In Section 7 we generalize these results to arbitrary order
and in section 8 we comment about our results in general
cosmological setting.

\section{The Density PDF from Convergence Maps}
Our formalism is based on a hierarchical {\em ansatz} for the
matter correlation functions. In principle the entire set of
$N$-point correlation functions must be computed by solving a
coupled series of non-linear integro-differential equations known
as BBGKY hierarchy. Unfortunately, no such solution exists at
present. On the other hand, the  Vlasov-Poisson system in the
fluid limit does admit hierarchical scaling solutions. Motivated
by this fact, several well-known hierarchical {\em ansatze} have
been suggested  which are very helpful in understanding the
non-linear dynamics of gravitational clustering. One of these {\em
ansatz} is particularly interesting, namely the one proposed by
Balian \& Schaeffer (1989) and developed considerably by
Bernardeau \& Schaeffer (1992). These studies are based on
multi-point cell count statistics and their scaling properties. To
apply it to lensing statistics is reasonably straightforward in
principle, but requires some preliminaries.

We adopt the following line element to describe the background
geometry:
\begin{equation}
d\tau^2 = -c^2 dt^2 + a^2(t)( d\chi^2 + r^2(\chi)d^2\Omega),
\end{equation}
where $a(t)$ is the expansion factor. The angular diameter
distance $r(\chi)= K^{-1/2}\sin [K^{-1/2} \chi]$ for positive
spatial curvature, $r(\chi) = (-K)^{-1/2}\sinh [(-K)^{-1/2}\chi] $
for negative curvature, and $r(\chi)=\chi$ for a flat universe. In
terms of $H_0$ and $\Omega_0$, $K= (\Omega_0 -1)H_0^2$.

We shall focus on the properties of the convergence $\kappa({\bf
\gamma})$ produced by lensing. In the standard terminology of
gravitational lensing this is simply the ratio of the surface mass
density $\Sigma({\bf \gamma})$ (observed in a direction ${\bf
\gamma}$ to the critical surface density for lensing $\Sigma_{\rm
cr}$. In our background geometry this reduces to a weighted
integral of the density fluctuation $\delta({\bf x})$ along the
line of sight:
\begin{equation}
\kappa({\bf \gamma}) = \inc {d\chi}
\omega(\chi)\delta\left[r(\chi),{\bf \gamma}\right].
\end{equation}
If all the sources are at the same redshift, one can express the
weight function $\omega(\chi) = 3/4a c^{-2}H_0^2 \Omega_m r(\chi)
r(\chi_s - \chi)/ r( \chi_s)$,  where $\chi_s$ is the comoving
radial distance to the source. (This approximation is not
essential to the calculation, and it is easy to modify what
follows for a more accurate description.) Using the definitions we
have introduced above we can compute the projected variance of
$\kappa$ in terms of the power spectrum of density fluctuations
$P({\bf l})$ using
\begin{equation}
\langle \kappa_s^2\rangle = \inc d {\chi_1} {\omega^2(\chi_1)
\over r^2(\chi_1)} \int {d^2 {\bf l} \over (2 \pi)^2}~P { \left(
{{\bf l}\over r(\chi)} \right)} W^2(l\theta_0),
\end{equation}
(Limber 1954). The higher order moments of the convergence field
can be written in the form:
\begin{eqnarray}
\langle \kappa^N_{s} \rangle = S_N \inc d {\chi} {\omega^N(\chi) \over
r^{2(N-1)}(\chi)} \kappa_{\theta_0}^{N-1},
\end{eqnarray}
(Munshi \& Coles 2000a), where the $S_N$ are defined by analogy
with the hierarchical parameters usually used in the context of
galaxy counts-in-cells.

If the mean number of  galaxies per cell is $\bar{N}$
 and the volume-average of the two-point correlation
function over the cell is
\begin{equation}
\bar{\xi}_2=\frac{1}{V^2}\int \int \xi_2({\bf r}_1, {\bf r}_2)
dV_1 dV_2,
\end{equation}
then the generalisation of equation the previous equation
\begin{equation} \bar{\xi}_N=\frac{1}{V^N} \int \ldots \int
\xi_N({\bf r}_1\ldots {\bf r_N}) dV_1\ldots dV_N. \label{eq:xibar}
\end{equation}
leads to the description of higher-order statistical properties of
galaxy counts in terms of the scaling parameters $S_N$ constructed
from the $\bar{\xi}_N$ via
\begin{equation}
S_N=\frac{\bar{\xi}_N}{\bar{\xi}_2^{N-1}}.\label{eq:sp}
\end{equation}

In equation (4) we have also introduced a new notation, namely
that
\begin{equation}
\kappa_{\theta_0} = \int { d^2 {\bf l} \over (2\pi)^2} P \left (
{l \over r(\chi)} \right ) W^2(l\theta_0),
\end{equation}
in order to take account of smoothing over an angular scale
$\theta_0$. The function $W(l\theta_0)$ is the window function for
the smoothing. In many studies a top-hat filter is used for
smoothing the convergence field, but our study can be extended to
compensated filters (Schneider et al. 1998; Reblinsky et al.
1999), which may be more appropriate for observational purposes.

In order to compute the PDF of the smoothed convergence field
$\kappa_s$, we will begin by constructing its associated cumulant
generating function $\Phi^{1+\kappa}(y)$
\begin{equation}
\Phi^{1 + {\kappa}}(y) = y + \sum_{p=2}^ {\infty} {{\langle
\kappa^p_s \rangle} \over \langle \kappa_s
\rangle^{p-1}} y^p.
\end{equation}
Using the expression for the higher moments of the convergence
field $\kappa_s$ we can write
\begin{equation}
\Phi^{1 + {\kappa}}(y) = y + \int_0^{\chi_s} \sum_{N=2}^{\infty}
{ (-1)^N \over N!} S_N { \omega^N(\chi) \over r^{2(N-1)} ( \chi)}
\var^{ (N-1)} { y^N \over {\langle \kappa_s^2 \rangle}^{(N-1)} }.
\end{equation}
We can now  express  $\Phi^{1 + \kappa}(y)$, in terms of the
cumulant generating function of the matter distribution, $\phi(y)$
using \begin{equation} \phi(y)=\sum_{p=1}^{\infty} \frac{S_P}{p!}
y^p,
\end{equation} in which the constants $S_p$ are the0
hierarchical parameters discussed above. In terms of
$\phi^{\eta}(y)$ we get
\begin{equation} \Phi^{1+\kappa}(y) =
\int_0^{\chi_s} r^2(\chi) d \chi \left [\kappa_{\theta_0} \over
\langle \kappa^2_{s} \rangle \right ]^{-1} \phi^{\eta} \Big [y
{\omega (\chi) \over r^2 (\chi)} {\var \over {\langle \kappa_s^2
\rangle}} \Big ] - y\int_0 ^{\chi_s} \omega(\chi) d \chi.
\label{valag}
\end{equation}
The second term in eq. (\ref{valag}) comes from the $N=1$ term in
the expansion of $\Phi^{1+\kappa}$. Note that we have used the
fully non-linear generating function $\phi^{\eta}$ for the
cumulants, though we will use it to construct a generating
function in the quasi-linear regime.

The analysis simplifies if we define a new reduced convergence
field $\eta_s$ defined by
\begin{equation} \eta_s = {{
\kappa_s - \kappa_{m} } \over -\kappa_{m}} = 1 + {\kappa_s \over
|\kappa_{m}| } ,
\end{equation}
where the minimum value of $\kappa_s$, denoted $\kappa_{m}$,
occurs when the line of sight goes through regions that are
completely devoid of matter (i.e. $\delta = -1$ all along the line
of sight):
\begin{equation}
\kappa_{m} = - \int_0^{\chi_s} d \chi \omega(\chi) .
\end{equation}

Although $\kappa({\theta_0})$ depends on the smoothing angle, its
minimum value $\kappa_{m}$ depends only on the source redshift and
background geometry of the universe and is independent of the
smoothing radius. In terms of the reduced convergence $\eta_s$,
the cumulant generating function is given by,
\begin{equation}
\Phi^{\eta} (y) = { 1 \over [\kmin]^2} \int_0^{\chi_s}r^2(\chi) d \chi
\Big [{ \var \over {\langle \kappa_s^2 \rangle} }\Big ]^{-1} \phi^{\eta} \Big [y\kmin
{\omega(\chi) \over r^2(\chi)}   {\var \over {\langle \kappa_s^2 \rangle}}   \Big ]
\end{equation}
The new cumulant generating function
 $\Phi^{\eta}(y)$ satisfies the normalization constraints $S_1 = S_2 = 1$.

The scaling function associated with $P^{\eta}(\eta)$ can now be
related to the matter scaling function $h(x)$ introduced in Munshi
et al. (1999a) in the context of matter clustering. This function
is defined in terms of the scaling variable $x=N/N_c$, where
$N_c=\bar{N}\bar{\xi}_2$. In the hierarchical {\em ansatz} the
probability distribution $P(N)$ can be expressed in scaled form as
\begin{equation}
P(N)=\frac{1}{\bar{\xi_2} N_c} h(x).
\end{equation}
The quantity $h(x)$ can then be expressed in terms of an integral
of the form
\begin{equation} h(x)  =  -{1 \over 2\pi
i}\int_{i\infty}^{i\infty} dy~y\sigma(y) \exp(yx).
\end{equation}
This can be extended to the bias factor of cells of occupancy $N$
via
\begin{equation}
P(N)b(N) = { 1 \over \xi_2 N_c} h(x)b(x)
\end{equation}
and so on for higher-order statistics; see Munshi et al. (1999a)
for more details.

In the present context we instead define
\begin{equation}
H^{\eta} (x) = - \int_{-\infty}^{\infty} { dy \over 2 \pi i} \exp
(x y) \Phi^{\eta} (y).
\end{equation}
Using this definition we can write
\begin{equation}
H^{\eta} (x) = { 1 \over \kappa_m }
\int_0^{\chi_s} \omega (\chi) d \chi \Big [ { r^2(\chi)\langle
\kappa_{s}^2\rangle \over \omega (\chi) \var \kappa_m} \Big ]^2
   h^{\eta} \left [{\langle
\kappa_{s}^2\rangle \over \var} {r^2(\chi) \over \omega (\chi)}{x
\over \kappa_m } \right ].
\end{equation}

These equations apply to the most general case within the small
angle approximation, but can be simplified considerably using
further approximations. In the following we will assume that the
contribution to the integrals involved in the expressions for
$\chi$ can be replaced by an average value coming from the maximum
of $\omega(\chi)$, i.e. $\chi_c$ ($0<\chi_c<\chi_s$). This idea
leads to the following approximate expressions:
\begin{eqnarray}
 |\kappa_{m}| & \approx & {1\over 2} \chi_s \omega(\chi_c)\nonumber \\
 \one & \approx & {1\over 2} \chi_s  \omega^2(\chi_c) \Big [ {d^2 k \over
(2\pi)^2} {\rm P(k)} W^2(k r(\chi_c) \theta_0) \Big ].
\end{eqnarray}
Using these approximations we can write
\begin{eqnarray}
\Phi^{\eta}(y) & = & \phi^{\eta}(y)\nonumber\\
 H^{\eta}(x) & = & h^{\eta}(x).
\end{eqnarray}
We thus find that the statistics of the smoothed underlying field
 and those of the  reduced
convergence $\eta$ are exactly the same in this approximation.
(The approximate functions  $\Phi^{\eta}$ and $h^{\eta}(x)$ do
satisfy the correct normalization constraints, but we omit the
details here.)

Although it is possible to integrate the exact expressions of the
scaling functions, there is some uncertainty involved in the
actual determination of these functions and their associated
parameters which must be inferred from numerical simulations. See
Colombi et al. (1996), Munshi et al. (1999a) and Valageas et al.
(1999) for a detailed description of the effect of finite volume
corrections involved in their estimation. In numerical studies
involving ray tracing simulations (Munshi \& Jain 1999b) it has
been found that the minimal hierarchical model of Bernardeau \&
Schaeffer (1992), which has  only one free parameter to be
determined  from numerical simulations, can reproduce results from
simulations very accurately.

Another useful approach is to construct an  Edgeworth expansion of
the PDF, starting with the simpler form $P(\eta_s)$. This can then
 be used to construct a series for $P(\kappa_s)$.
The Edgeworth expansion (Bernardeau \& Kofman 1994) is meaningful
when the variance is less than unity, a condition that guarantees
a convergent series expansion in terms of Hermite polynomials
$H_n(\nu)$, of order n and with $\nu = \eta_s
/\sqrt{(\xi_{\eta_s}}$).The relevant series expansion is
\begin{equation} P^{\eta}(\eta_s) = { 1 \over  \sqrt 2 \pi
{\bar \xi}_{\eta_s} } \exp ( - {\nu^2 \over 2} ) \Big [ 1 + {\sqrt
{\bar \xi}_{\eta}} {S_3^{\eta} \over 6} H_3(\nu)+ {\sqrt {\bar
\xi}_{\eta}}^2 \Big ( {{S_4^{\eta} \over 24} H_4(\nu)+
{{S_3^{\eta}}^2 \over 72}} H_6(\nu) \Big ) + \dots  \Big ]
\end{equation}
This is usually applied to a quasi-linear analysis of matter
clustering  using perturbative expressions for the cumulants.
However, the $S^{\eta}_N$ parameters needed in the expansion of
$P(\eta)$ are from the highly non-linear regime. Although the
variance is still smaller than unity, the parameters that
characterize it emerge from the highly non-linear dynamics of the
underlying dark matter distribution.

The magnification $\mu$ can also be used instead of $\kappa$
according to the weak lensing relation, $\mu_s = 1 + 2 \kappa_s$.
Its minimum value can be related to $\kappa_{m}$ defined earlier
via $\mu_{m} = 1 + 2 \kappa_{m}$. Finally, the reduced convergence
$\eta$ and the magnification $\mu$ can be related by the following
equation:
\begin{equation}
\eta_s = {\mu_s - \mu_{m} \over 1 - \mu_{m}}
\end{equation}
(Valageas 1999). We can now express the relations connecting the
probability distribution function for the smoothed convergence
statistics $\kappa_s$, the reduced convergence $\eta_s$ and the
magnification $\mu_s$ as,
\begin{equation}
P^{\kappa}(\kappa_s) = 2 P^{\mu}(\mu_s) = P^{\eta}(\eta_s) { 2
\over ( 1 -  \mu_{m}) } = P^{\eta}(\eta_s) { 1 \over
|\kappa_{m}|}.
\end{equation}

The formalism which we have developed for one-point statistics
such as the PDF and the VPF can also be extended to compute the
bias and higher order cumulants associated with spots in $\kappa$
maps above a certain threshold. The statistics of such spots can
be associated with the statistics of over-dense regions in the
underlying mass distribution representing the collapsed objects. A
detailed analysis of these issues will be presented elsewhere
(Munshi \& Coles 2000b).

\section{Bias of Collapsed Objects From Convergence Maps}

In order To compute the bias associated with the peaks in the
convergence field we must first develop an analytic expression for
the joint generating function $\beta(y_1, y_2)$ for the
convergence field $\kappa_s$. For that we will use the usual
definition for the two-point cumulant correlator $C_{pq}$ of the
convergence field
\begin{equation}
C^{\kappa}_{pq} = {\langle \kappa_s(\gamma_1)^p
\kappa_s(\gamma_2)^q \rangle \over \langle k_s^2 \rangle^{p+q-2}
\langle \kappa_s(\gamma_1) \kappa_s(\gamma_2) \rangle } =
C^{\kappa}_{p1} C^{\kappa}_{q1}.
\end{equation}
For a complete treatment of lower order moments of $\kappa_s$, see
Munshi \& Coles (2000a). Like its counterpart for the density
field, the two-point generating function of the convergence field
can  be expressed as a product of two one-point generating
functions:
\begin{equation}
{}_2\beta_{\kappa}(y_1, y_2) = \sum_{p,q}^{\infty}
{C^{\kappa}_{pq} \over p! q!} y_1^p y_2^q = \sum_{p}^{\infty}
{C^{\kappa}_{p1} \over p!} y_1^p \sum_{q}^{\infty}
{C^{\kappa}_{q1} \over q!} y_2^q  = \beta_{\kappa}(y_1)
\beta_{\kappa}(y_2)\equiv \tau_{\kappa}(y_1) \tau _{\kappa}(y_2),
\end{equation}
where the function $\tau_{\kappa}$ will be used later. The
factorization property of the generating function depends on the
factorization property of the cumulant correlators: $C^{\eta}_{pq}
= C^{\eta}_{p1} C^{\eta}_{q1}$. Such a factorization is possible
when the correlation between two patches in the directions
$\gamma_1$ and $\gamma_2$ is smaller than the variance on the
scale of one patch. Using this and starting with
\begin{equation}
{}_2\beta_{\kappa}(y_1, y_2) = \sum_{p,q}^{\infty} {1 \over p! q!}
{ y_1^p y_2^q\over \langle \kappa_s^2 \rangle^{p+q-2} } {\langle
\kappa_s(\gamma_1)^p \kappa_s(\gamma_2)^q \rangle  \over \langle
\kappa_s(\gamma_1) \kappa_s(\gamma_2)\rangle },
\end{equation}
and then inserting the integral expression for the cumulant
correlators in the hierarchical {\em ansatz} we obtain
\begin{eqnarray}
{}_2\beta_{\kappa}(y_1, y_2) &=& \sum_{p,q}^{\infty} {C^{\eta}_{pq} \over p! q! } { 1 \over
\langle \kappa_s^2 \rangle^{p+q -2}} { 1 \over \langle
\kappa_s(\gamma_1) \kappa_s (\gamma_2) \rangle } \int_0^{\chi_s} d\chi
{ \omega^{p+q} \over r^{2(p+q -1)} } \corr \var^{p+q-2} y_1^p y_2^q.
\end{eqnarray}
This expression involves the definition
\begin{equation}
\kappa_{\theta_{12}} = \int { d^2 {\bf l} \over (2\pi)^2} P \left
( {l \over r(\chi)} \right ) W^2(l\theta_0) \exp(il \theta_{12}).
\end{equation}

\begin{figure}
\protect\centerline{
 \epsfysize = 1.5truein
 \epsfbox[4 4 377 194]
 {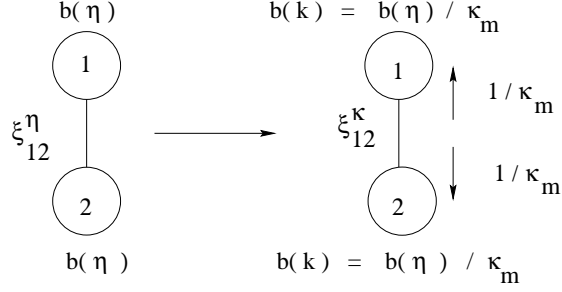} }
 \caption{Construction of bias function $b(\kappa_s)$ for the smoothed convergence
 field from the bias $b(\eta_s)$ associated with the smoothed reduced
 convergence field $\eta_s$.}
\end{figure}

It is possible to simplify this approximation further by grouping
the summations over dummy variables $p$ and $q$. This is useful
 to establish the factorization property of the
two-point (joint) generating function for bias ${}_2\beta(y_1,
y_2)$. Note that
\begin{eqnarray}
{}_2\beta_{\kappa}(y_1, y_2) &=& \inc r^2(\chi)  d\chi  \left [ {\corr
 \over \two } \right ] \left [ { \var \over \one} \right ]^{-2} \sum_{pq}^{\infty}
 {C^{\eta}_{pq} \over p! q!} \left [ y_1{\omega(\chi) \over r^2
(\chi)} { \var \over \one}  \right ]^p \left [y_2{\omega(\chi)
\over r^2 (\chi)} { \var \over \one} \right ]^q.
\end{eqnarray}
We can now decompose the double sum over the two indices into two
separate sums over individual indices. Finally using the
definition of the one-point generating function for the cumulant
correlators we can write:
\begin{eqnarray}
{}_2\beta_{\kappa}(y_1, y_2) &=& \inc r^2(\chi)  d\chi  \left [ { \corr \over
 \two }\right ] \left [ { \var
\over \one
} \right ]^{-2} \tau^{\eta} \left [ y_1{\omega(\chi) \over r^2
(\chi)} { \var \over \one} \right ]  \tau^{\eta} \left [y_2{\omega(\chi) \over r^2
(\chi)} { \var \over \one}  \right ]
\end{eqnarray}
The above expression is quite general. It depends on the
small-angle and large separation approximations, but is valid for
any particular model for the tree-correlation hierarchy. However
it can be seen that the projection effects encoded in the
line-of-sight integration do not allow us to express  the
two-point generating function $\beta_{\eta}(y_1, y_2)$ simply as a
product of two one-point generating functions $\beta^{\eta}(y)$ as
can be done in the case of the underlying density field.

As in the case of the derivation of the probability distribution
function for the smoothed convergence field $\kappa_s$, matters
simplfy considerably if we use the reduced smoothed convergence
field $\eta_s$. We have already shown that the PDF associated with
$\eta_s$ and the underlying mass distribution are the same under
certain approximations. An identical ressult result indeed holds
good for higher order statistics, including the bias. The
cosmological dependence of the statistics of $\kappa_s$ field is
encoded in $k_{m}$ and the choice of the new variable $\eta$
renders its related statistics almost independent of background
cosmology. Repeating the above analysis again for the $\eta_s$
field now we can express the cumulant correlator generating
function for the reduced convergence field $\eta_s$ as follows:
\begin{eqnarray}
{}_2\beta_{\eta}(y_1, y_2) = { 1 \over [\kappa_m]^2}\inc r^2(\chi)
d\chi \left [ { \corr \over \two } \right ] \left [ { \var
\over \one
} \right ]^{-2} \tau^{\eta} \left [ y_1 \kappa_m {\omega(\chi) \over r^2
(\chi)}{ \var \over \one} \right ]  \tau^{\eta} \left [ y_2 \kappa_m {\omega(\chi) \over r^2
(\chi)}{ \var \over \one} \right ].
\end{eqnarray}

Although the above expression is very accurate and describes an
important relationship between the density field and the
convergence field, it is difficult to use for any practical
purpose. It is also important to note that scaling functions such
as $h(x)$ for the density probability distribution function and
$b(x)$ for the bias associated with over-dense objects are
typically estimated from numerical simulations, especially in the
highly non-linear regime. Such estimations are subject to many
uncertainties, such as the finite size of the simulation box. It
has been noted in earlier studies that such uncertainties lead to
large errors in estimates of $h(x)$. Estimates of the scaling
function associated with the bias  $b(x)$ is even more
complicated, owing to the fact that these quantities are even
worse affected finite size of the catalogues. It is consequently
not fruitful to  integrate the exact integral expression we have
derived above. Instead we will replace all line-of-sight integrals
with approximate values. An  exactly similar approximation was
used by Munshi \& Jain (1999a) to simplify the one-point
probability distribution function for $\kappa_s$,  and it was
found to be in good agreement with numerical simulations. In this
approximation
\begin{equation}
\two  \approx {1\over 2} \chi_s  \omega^2(\chi_c) \Big [ {d^2 k
\over (2\pi)^2} {\rm P(k)} W^2(k r(\chi_c) \theta_0) \exp [i k
r(\chi_c) \theta_{12}] \Big ].
\end{equation}
This gives us the leading order contributions to the integrals
needed and we can also check that, to this order, we recover the
factorization property of the generating function, i.e.
\begin{equation}
{}_2\beta_{\eta}(y_1, y_2) = \tau^{\eta}(y_1) \tau^{\eta}(y_2) =
\tau_{1+\delta}(y_1) \tau_{1+\delta}(y_2) \equiv
\tau(y_1)\tau(y_2).
\end{equation}
It is also interesting to note that $C_{p1}^{\kappa} =
C_{p1}^{\eta} {\kappa_m}^{p-1}$.

In this level of approximation, the factorization property of the
cumulant correlators  means that the  bias function $b(x)$
associated with the peaks above a given threshold in the
convergence field $\kappa_s$ also obey a factorization property,
just as is the case in the density field counterpart (e.g. Coles,
Munshi \& Melott 2000):
\begin{equation}
b^{\eta}(x_1) h^{\eta}(x_1) b^{\eta} (x_2) h^{\eta} (x_2) =
 b_{1 + \delta}(x_1) h_{1 + \delta}(x_1) b_{1+\delta} (x_2) h_{1+\delta}
 (x_2),
\end{equation}
in which we  have used the relation between $\beta(y)$ and $b(x)$
given above:
\begin{equation}
b^{\eta}(x) h^{\eta}(x) = -{ 1 \over 2 \pi i}
\int_{-i\infty}^{i\infty} dy \tau (y) \exp (xy) \label{ber1}.
\end{equation}
We established a similar correspondence between the convergence
field and density field in the case of one-point probability
distribution function  in the previous section.

The differential bias, i.e. the bias associated with a particular
overdensity, is much more difficult to measure from numerical
simulations than its integral counterpart. From now on we
therefore concentrate on the bias associated with peaks above
certain threshold, which can be expressed in a similar form to
that given above:
\begin{equation}
b^{\eta}(>x) h^{\eta}(>x) = -{ 1 \over 2 \pi i}
\int_{-i\infty}^{i\infty} dy {\tau (y)\over y} \exp (xy).
\label{ber2}
\end{equation}
(Munshi et al. 1999a). Although the bias $b(x)$ associated with
the convergence field and that related to the underlying density
field are exactly equal, the variance associated with the density
field is very much higher than that of the convergence field.
Projection effects bring down the latter to less than unity,
meaning that we have to use the integral definition of bias to
recover it from its generating function (see eq.(\ref{ber1}) and
eq.(\ref{ber2})).

We can now write down the full two-point probability distribution
function for two correlated spots in terms of the convergence
field $\kappa$ and its reduced version $\eta$:
\begin{eqnarray}
p^{\kappa}(\kappa_1, \kappa_2)d\kappa_1 d\kappa_2 & = &
p^{\kappa}(\kappa_1) p^{\kappa}(\kappa_2)\left[ 1 +
b^{\kappa}(\kappa_1) \xi^{\kappa}_{12} b^{\kappa}(\kappa_2)\right]
d\kappa_1 d\kappa_2 \nonumber \\ p^{\eta}(\eta_1, \eta_2)d\eta_1
d\eta_2 & = & p^{\eta}(\eta_1) p^{\eta}(\eta_2)\left[ 1 +
b^{\eta}(\eta_1) \xi^{\eta}_{12} b^{\eta}(\eta_2)\right] d\eta_1
d\eta_2.
\end{eqnarray}
Following from the analysis presented in the previous section, we
note that $p(\kappa_s) = p(\eta_s)/k_{m}$ and that
$\xi^{\kappa}_{12} = \xi^{\eta}_{12}/ \kappa_{m}^2$. Using these
relations we can now write
\begin{equation}
b^{\kappa}(\kappa  ) = {b^{\eta}(\eta  ) \over {k_{m}}}.
\end{equation}
This is one of the main results of this analysis  and has already
been shown by Munshi (2000) to be in very good agreement
with numerical ray tracing simulations.

\section{Three-Point Statistics of Collapsed Objects From Convergence Maps}

A non-linear {\em ansatz} for higher-order correlations can  be
used not only to compute the correlation hierarchy for the
underlying mass distribution but also the multi-point statistics
of overdense cells, which represent  collapsed objects (Bernardeau
\& Schaeffer 1992, 1999; Munshi et al. 199a,b,c; Coles et al.
1999). Whereas such studies generally concentrate mainly on
three-dimensional statistical properties our aim in this paper is
to investigate how much one can learn from such {\em ansatze}
about the projected density field obtained from weak lensing
surveys. In  earlier sections we showed that the bias associated
with peaks in projected density field can be very accurately
modelled. We will now extend these results using multi-point
cumulant correlators to show that such an analysis can also be
performed for the skewness of overdense objects in the projected
density field.

\begin{figure}
\protect\centerline{
 \epsfysize = 1.5truein
 \epsfbox[172 9 896 202]
 {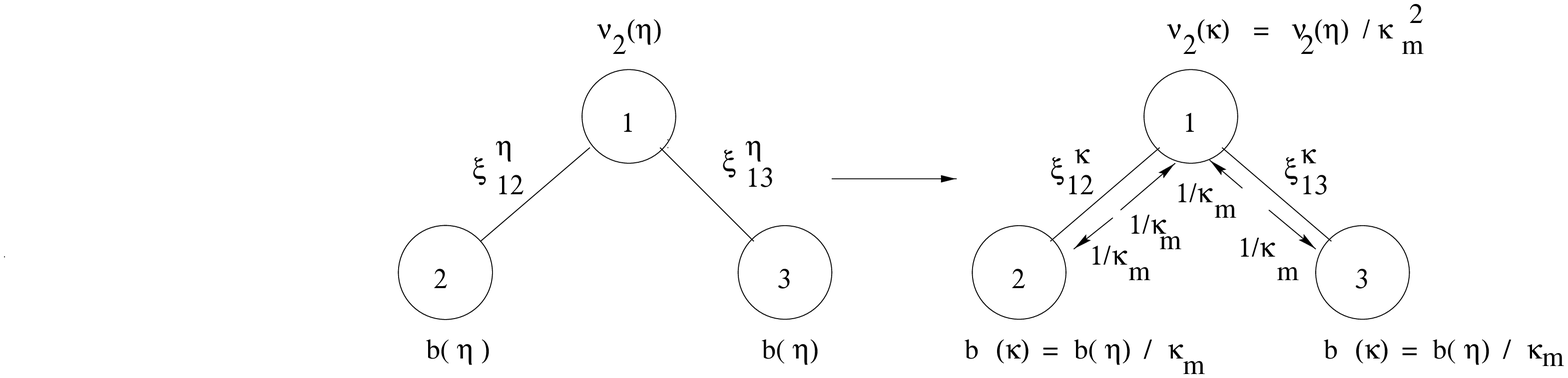} }
\caption{Construction of the bias function $b(\kappa_s)$ for the
smoothed convergence
 field from the bias $b(\eta_s)$ associated with the smoothed reduced
 convergence field $\eta_s$.}
\end{figure}

The bias of collapsed objects is connected with the two-point
cumulant correlators of the underlying mass distribution. In a
very similar fashion, the three-point cumulant correlators of
underlying mass distribution can be related to the skewness of
collapsed objects. The three-point cumulant correlators for
projected mass distribution, smoothed with an angle $\theta_0$ can
be written as (Munshi \& Coles 2000a)
\begin{eqnarray}
\langle \kappa_s(\gamma_1)^p \kappa_s (\gamma_2)^q \kappa_s(\gamma_3)^r
\rangle_c = \int_0^{\chi_s} d \chi { \omega^{p+q+r} \over r^{2(p+q+r-1)} }
\left [ C_{p1}^{\eta} \corr  C_{q11}^{\eta}  \corri C_{r1}^{\eta} \var^{p+q+r-3}
+ {~~~~\rm cyc. perm.} \right ].
\end{eqnarray}
The three-point cumulant correlators for the projected density
field can be defined by the following equation:
\begin{eqnarray}
C_{pqr}^{\kappa} & = & {\langle \kappa_s(\gamma_1)^p \kappa_s
(\gamma_2)^q \kappa_s(\gamma_3)^r \rangle_c  \over \langle
\kappa_s(\gamma_1) \kappa_s (\gamma_2)\rangle_c \langle
\kappa_s(\gamma_2) \kappa_s (\gamma_3) \rangle_c \langle
\kappa^2(\gamma) \rangle_c^{p+q+r-3}}\nonumber\\
 & = & \int_0^{\chi_s} d \chi { \omega^{p+q+r} \over r^{2(p+q+r-1)} }
C_{p1}^{\eta} {\corr \over \langle \kappa_s(\gamma_1)
\kappa_s(\gamma_2) \rangle_c } C_{q11}^{\eta} { \corri \over
\langle \kappa_s(\gamma_2) \kappa_s(\gamma_3) \rangle_c }
 C_{r1}^{\eta} { \var^{(p+q+r-3)}
\over \langle \kappa_s^2(\gamma) \rangle_c^{(p+q+r-3)}}.
\end{eqnarray}
In what follows we present only one representative term for each
topology at each order; other terms can be obtained by cyclic
permutation of indices. This simplifies the presentation  because
the result of the final analysis can also be achieved by simple
cyclic permutations of indices. Note that, because of the
line-of-sight integration we can not decompose the generating
function as a product of three generating functions of single
variables.

Introducing the new variable $\eta$ as before we find that
\begin{eqnarray}
C_{pqr}^{\kappa} & = & {\langle \eta_s(\gamma_1)^p \eta_s
(\gamma_2)^q \eta_s(\gamma_3)^r\rangle_c \over \langle
\eta_s(\gamma_1) \eta_s (\gamma_2) \rangle_c \langle
\eta_s(\gamma_2) \eta_s (\gamma_3) \rangle_c \langle
\eta^2(\gamma) \rangle_c^{p+q+r-3}} \nonumber \\ & = &
{(\kappa_{m})}^{p+q+r-2}{\langle \kappa_s(\gamma_1)^p \kappa_s
(\gamma_2)^q \kappa_s(\gamma_3)^r\rangle_c \over \langle
\kappa_s(\gamma_1) \kappa_s (\gamma_2)\rangle_c \langle
\kappa_s(\gamma_2) \kappa_s (\gamma_3)\rangle_c \langle
\kappa^2(\gamma)
\rangle_c^{p+q+r-3}}\nonumber\\
 & = & (\kappa_{m})^{p+q+r-2} C_{pqr}^{\eta} \end{eqnarray}
The generating function for the convergence field $\kappa_s$ can
then be written as a summation over different indices:
\begin{eqnarray}
{}_3\beta_{\kappa}(y_1,y_2,y_3) & = & \sum_{p=1}^{\infty}
C^{\eta}_{pqr} {y_1^p \over p!} {y_2^q \over q!} {y_3^r \over
 r!} \nonumber\\
 & = &  \int_0^{\chi_s}{r^2(\chi) \over \omega^2(\chi) }  d
\chi \sum _{p=1}^{\infty} {
(-\kappa_{m} )^{p-1} \over p!}C_{p1}^{\eta} { \var^p \over
 \one^p} \times {\corr \over \two}
\sum_{q=1}^{\infty}{ (-\kappa)^q \over q!}  C_{q11}^{\eta} {
\var^q \over \one^q} {\corri \over \two} \nonumber \\ & & \times
\sum_{r = 1}^{\infty} { (-\kappa_{m})^{r-1} \over r!}
C_{r1}^{\eta} {\var^r \over \one^{2r} } y_3^r \left [ \var \over
\one \right ]^{-3}.
\end{eqnarray}
Making a change of variables from convergence $\kappa_s$ to
reduced convergence $\eta_s$ we can write
\begin{eqnarray}
{}_3\beta_{\eta}(y_1,y_2,y_3) & = & { 1 \over [\kmin]^2}
\int_0^{\chi_s} d \chi \left[ {\corr \over \two} \right ]\nonumber
\left [ {\corri \over \two }  \right ]
 \left [{\var \over \one } \right ]^{-3} \nonumber \\ & & \times~
\mu_1^{\eta} \left [ y_1 \kmin { {\omega(\chi) \over r^2(\chi)} }{\var \over
\one} \right ]
 \mu_2^{\eta} \left [y_2 \kmin { {\omega(\chi) \over r^2(\chi)} }{\var \over
\one} \right ]
 \mu_1^{\eta} \left [y_3 \kmin { {\omega(\chi) \over r^2(\chi)} }{\var \over
\one} \right ].
\end{eqnarray}

As we mentioned above, it is  possible to integrate the above
equation (which is derived using only small angle approximation).
However, note that due to  line-of-sight effects it is no longer
possible to separate the cumulant correlators of the convergence
field in the same way as can be done for the underlying mass
distribution. Using the approximations for the expressions which
we have already used to simplify the generating function for
two-point cumulant correlators,  i.e.
\begin{equation}
\langle \kappa_s(\gamma_i) \kappa_s(\gamma_j)  \rangle  \approx
{1\over 2} \chi_s  \omega^2(\chi_c) \Big [ {d^2 k \over (2\pi)^2}
{\rm P(k)} W^2(k r(\chi_c) \theta_0) \exp [i k r(\chi_c)
\theta_{ij}] \Big ],
\end{equation}
we can write
\begin{equation}
{}_3\beta_{\eta}(y_1,y_2,y_3) = \mu_1^{\eta}(y_1)\mu_2^{\eta}(y_2)\mu_1^{\eta}(y_3).
\end{equation}
So at this level of approximation we again obtain a factorization
property of the cumulant correlators for the reduced convergence
field. Also notice that such an approximation would man that the
third order reduced cumulant correlators for convergence field can
be related to the corresponding quantity for the underlying mass
distribution by
\begin{equation}
C_{p11}^{\eta} = C_{p11}^{\kappa} (-\kappa_m)^p.
\end{equation}
The three-point joint PDF for the convergence field and the
reduced convergence field now can be expressed as
\begin{eqnarray}
p^{\kappa}(\kappa_1, \kappa_2, \kappa_3) & = &
p^{\kappa}(\kappa_1)p^{\kappa}(\kappa_2)p^{\kappa}(\kappa_3)
\left[ 1 + b^{\kappa}(\kappa_1)\xi^{\kappa}_{12}
b^{\kappa}(\kappa_2) + b^{\kappa}(\kappa_1)\xi^{\kappa}_{12}
\nu_2^{\kappa}(\kappa_2) \xi^{\kappa}_{23} b^{\kappa}(\kappa_3) +
{\rm cyc. perm.} \right] \nonumber \\ p^{\eta}(\eta_1, \eta_2,
\eta_3) & = & p^{\eta}(\eta_1)p^{\eta}(\eta_2)p^{\eta}(\eta_3)
\left[ 1 + b^{\eta}(\eta_1)\xi^{\eta}_{12} b^{\eta}(\eta_2) +
b^{\eta}(\eta_1)\xi^{\eta}_{12} \nu_2^{\eta}(\eta_2)
\xi^{\eta}_{23} b^{\eta}(\eta_3) + {\rm cyc. perm.} \right]
\end{eqnarray}
The generating function for the reduced cumulant correlator
$C_{p11}^{\eta}$, i.e. $\mu_2^{\eta}(y)$ and its associated
scaling function $\nu_2^{\eta}(x)$ (see Munshi et al. 1999a,b) can
be related by
\begin{equation}
\nu_2^{\eta}(x) h^{\eta}(x) = -{ 1 \over 2 \pi i}
\int_{-i\infty}^{i\infty} dy \mu_2^{\eta} (y) \exp (xy),
\end{equation}
where the scaling variable $x$ has the same meaning as defined
earlier. One can also derive a similar expression for cumulative
$\nu_2^{\eta}(>x)$, i.e.  $\nu_2^{\eta}(>x)$ beyond a certain
threshold (see Munshi, Coles, Melott 1999a,b; Munshi, Melott \&
Coles 2000; Munshi et al. 1999; Munshi \& Melott 1998 for
details). Using the fact that
 $p(\kappa) = {p(\eta)/ \kappa_{m}}$;
$b(\kappa) = {b(\eta)/\kappa_{m}}$ and $\xi_{\kappa} =
\xi_{\kappa}/ \kappa_m^2$ we can finally obtain
\begin{equation}
\nu_2^{\kappa}(\kappa) = {\nu_2^{\eta}(\eta) \over {\kappa_m}^2}.
\end{equation}
Since the scaling function $\nu_2^{\kappa}(>x)$ encodes
information concerning the  skewness of collapsed objects beyond a
certain threshold or hot-spots in convergence maps it is
interesting to notice how such a quantity is directly related to
the three-point cumulant correlators of the background convergence
field. We can write the skewness $S_3^{\kappa}(>\kappa_s)$ of
those spots in convergence maps which cross certain threshold
$\kappa_s$ as
\begin{equation}
S_3^{\kappa}(>\kappa_s) = 3{\nu_2^{\kappa}(> \kappa_s) \over
b^{\kappa}(> \kappa_s)^2} = 3{\nu_2^{\eta}(>\eta_s)\over
b^{\eta}(> \eta_s)^2} = S_3^{\eta}(>\eta_s).
\end{equation}
It is easy to notice that, independent of cosmology,  spots which
cross certain threshold $\eta_s$ in reduced smoothed convergence
will have exactly same skewness as the skewness of the overdense
regions of the mass distribution $1+ \delta = {\rho_s \over
\rho_0}$ crossing the same threshold. This is one of the main
results in our analysis, in next sections we will show that this
results holds good to higher order. Indeed, it turns out that even
the  PDF of collapsed objects and ``hot-spots'' in reduced
convergence maps will be exactly equal.

\section{Four-Point Statistics of Collapsed Objects from Convergence Maps}

Four-point cumulant correlators can be analyzed in an exactly
similar manner as in the previous section, except that in this
case there are two distinct topologies (``snake'' and ``star''
topologies) which make contributions to the correlations. The star
topology possesses a completely new vertex of third order, the
snake topology (although of same order) is made of lower order
tree vertices. Therefore it also provides a unique consistency
check of the lower-order diagrams.

\begin{figure}
\protect\centerline{
 \epsfysize = 2.truein
 \epsfbox[2 2 705 308]
 {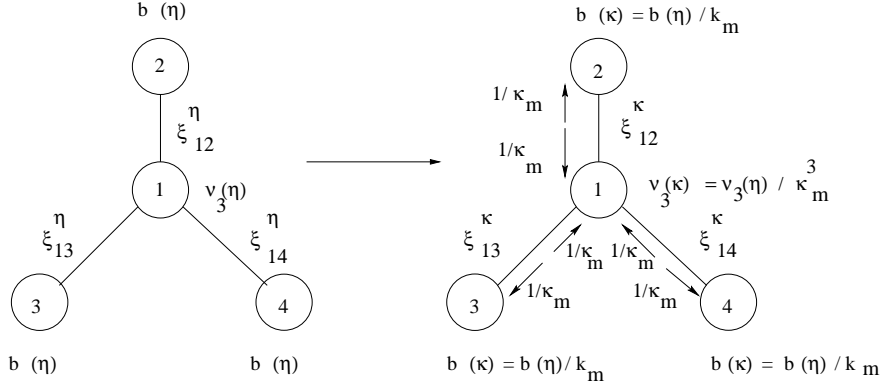} }
 \caption{Generating the third order vertex $\nu_3^{\kappa}(\kappa)$
 from its reduced convergence counterpart $\nu_3^{\eta}(\eta)$. Notice that
every link in the $\eta$ diagram  corresponds to an extra factor
of $1 / \kappa_m^2$ in the $\kappa$ diagram and hence one of these
$1 /\kappa_m$ gets attached to one of the two vertices connected
by such links. Such a simplification however is possible only
after some approximations, as described in the text. }
\end{figure}

The most general four-point cumulant correlator of arbitrary order
can be expressed as (Munshi, Melott \& Coles 1999a):
\begin{eqnarray}
\langle \eta^p_s(\gamma_1) \eta^q_s(\gamma_2) \eta^r_s(\gamma_3)
\eta^s_s(\gamma_4) \rangle_c & = & \big [ C_{p1}^\eta \langle
\eta_s(\gamma_1) \eta_s(\gamma_2) \rangle_c C_{q11}^{\eta} \langle
\eta_s(\gamma_2) \eta_s(\gamma_3) \rangle_c C_{r11}^{\eta} \langle
\eta_s(\gamma_3)\eta_s(\gamma_4) \rangle_c C_{s1}^{\eta}
 + {\rm cyc. perm.} + \dots \nonumber \\ &&+ { C_{p1}^\eta \langle \eta_s(\gamma_1)
\eta_s(\gamma_2) \rangle_c C_{q1}^{\eta} \langle \eta_s(\gamma_1)
\eta_s(\gamma_3) \rangle_c C_{r1}^{\eta}  \langle
\eta_s(\gamma_1)\eta_s(\gamma_4) \rangle_c C_{s111}^{\eta}} + {\rm
cyc. perm.}\big ] \nonumber\\ & & \times \langle \eta_s^2
\rangle^{(p+q+r+s-4)}
\end{eqnarray}
We will use the same superscript $\eta$ to denote both the reduced
convergence field and the underlying mass distribution. Both
fields display similar statistical properties, at least to the
level of approximation used here. Developing the four-point
function further yields
\begin{eqnarray}
\langle \eta^p_s(\gamma_1) \eta^q_s(\gamma_2) \eta^r_s(\gamma_3)
\eta^s_s(\gamma_4) \rangle_c & = & \int_0^{\chi_s} d \chi {
\omega(\chi)^{p+q+r+s} \over r^{2(p+q+r+s-1)}} \big [
C^{\eta}_{p1} \corr  C^{\eta}_{q11} \corri C^{\eta}_{r11} \corrj
C^{\eta}_{s1} + {\rm cyc. perm.} \nonumber\\&& + C^{\eta}_{p1}
{\kappa_{\theta}}_{12}  C^{\eta}_{q11}
 {\kappa_{\theta}}_{13}\  C^{\eta}_{r11}  {\kappa_{\theta}}_{14} C^{\eta}_{s1} + {\rm cyc. perm.} \big ]
 \var^{p+q+r+s-4}.
\end{eqnarray}
The snake contribution to four-point cumulant correlator can now
be written
\begin{eqnarray}
C_{pqrs}^{\rm snake}   &  =  &  {\langle \kappa_s^p(\gamma_1)
\kappa_s^q(\gamma_2) \kappa_s^r(\gamma_3)
\kappa_s^s(\gamma_4)\rangle_c^{\rm snake} \over \langle
\kappa_s(\gamma_1)\kappa_s(\gamma_2) \rangle \langle
\kappa_s(\gamma_3) \kappa_s(\gamma_4)\rangle \langle
\kappa_s^2(\gamma)\rangle^{p+q+r+s-4} }\nonumber\\ & = &
\int_0^{\chi_s} d \chi { \omega(\chi)^{p+q+r+s} \over
r^{2(p+q+r+s-1)}} C^{\eta}_{p1} \left [ {\corr \over \langle
\kappa_s(\gamma_1) \kappa_s(\gamma_2)\rangle}\right ]
C^{\eta}_{q11} \left [ {\corri \over \langle \kappa_s(\gamma_2)
\kappa_s(\gamma_3) \rangle } \right ] \  C^{\eta}_{r11}
\left [ {\corrj \over \langle \kappa_s(\gamma_3)
\kappa_s(\gamma_4) \rangle } \right ] C^{\eta}_{s1} \left [ {\var
\over \one}\right ]^{p+q+r+s-4}.
\end{eqnarray}
Considering the generating function for the snake terms, we obtain
the following expression:
\begin{eqnarray}
 {}_4\beta_{\kappa}^{\rm snake}(y_1,y_2,y_3,y_4)
& = & \sum_{pqrs}C_{pqrs}^{\rm snake}{y_1^p \over p!} {y_2^q \over
q!} {y_3^r \over r!} {y_4^s \over s!}\nonumber\\ & = & { 1 \over
\left [\kappa_m \right ] ^2} \int_0^{\chi_s} d \chi \left [ {\corr
\over \langle \kappa_s(\gamma_1)\kappa_s(\gamma_2) \rangle }
\right ] \left [ {\corri \over \langle
\kappa_s(\gamma_2)\kappa_s(\gamma_3) \rangle } \right ] \left [ {
\corrj \over \langle \kappa_s(\gamma_3)\kappa_s(\gamma_4) \rangle}
\right ]\nonumber\\ && \times  \mu_1^{\eta} \left [ {y_1 } {
\omega(\chi) \over r^2(\chi) } {\var \over \one} \right ]
\mu_2^{\eta} \left [ {y_2 } { \omega(\chi) \over r^2(\chi) } {\var
\over \one} \right ] \mu_2^{\eta} \left [ {y_3 } { \omega(\chi)
\over r^2(\chi) } {\var \over \one} \right ]\mu_1^{\eta} \left [
{y_4 } { \omega(\chi) \over r^2(\chi) } {\var \over \one} \right ]
\left [{\var \over \one} \right]^{-4}.
\end{eqnarray}
Once again, it is not possible to factorize the generating
function because of line-of-sight contributions. After changing
variables from $\kappa$ to $\eta$ we can write
\begin{eqnarray}
 {}_4\beta_{\eta}^{\rm snake}(y_1,y_2,y_3,y_4) & = &
\sum_{pqrs}C_{pqrs}^{\rm snake}{y_1^p \over p!} {y_2^q \over q!}
{y_3^r \over r!} {y_4^s \over s!} \nonumber\\ & = & { 1 \over
\left [\kappa_m \right ] ^2} \int_0^{\chi_s} d \chi \left [ {\corr
\over \langle \kappa_s(\gamma_1)\kappa_s(\gamma_2) \rangle }
\right ] \left [ {\corri \over \langle
\kappa_s(\gamma_2)\kappa_s(\gamma_3) \rangle } \right ] \left [ {
\corrj \over \langle \kappa_s(\gamma_3)\kappa_s(\gamma_4) \rangle}
\right ]\nonumber
\\&& \times  \mu_1^{\eta} \left [\kappa_m {y_1 } { \omega(\chi) \over
r^2(\chi) } {\var \over \one} \right ] \mu_2^{\eta} \left
[\kappa_m {y_2 } { \omega(\chi) \over r^2(\chi) } {\var \over
\one} \right ] \nonumber \\&& \times \mu_2^{\eta} \left [\kappa_m
{y_3 } { \omega(\chi) \over r^2(\chi) } {\var \over \one} \right
]\mu_1^{\eta} \left [\kappa_m {y_4 } {\omega(\chi) \over r^2(\chi)
} {\var \over \one} \right ] \left [{\var \over \one}
\right]^{-4}.
\end{eqnarray}
Similarly, for terms with star topologies, we can write
\begin{eqnarray}
C_{pqrs}^{\rm star}  &   = & {\langle \kappa_s^p(\gamma_1)
\kappa_s^q(\gamma_2) \kappa_s^r(\gamma_3)
\kappa_s^s(\gamma_4)\rangle_c^{\rm snake} \over \langle
\kappa_s(\gamma_1)\kappa_s(\gamma_2) \rangle \langle
\kappa_s(\gamma_3) \kappa_s(\gamma_4)\rangle \langle
\kappa_s^2(\gamma)\rangle^{p+q+r+s-4} }  \nonumber \\& = &
\int_0^{\chi_s} d \chi { \omega(\chi)^{p+q+r+s} \over
r^{2(p+q+r+s-1)}} C^{\eta}_{p111} \left [ {\corr \over \langle
\kappa_s(\gamma_1) \kappa_s(\gamma_2)\rangle}\right ]
C^{\eta}_{q1} \left [ {\kappa_{\theta_{13}}\over \langle
\kappa_s(\gamma_1) \kappa_s(\gamma_3) \rangle } \right ]
C^{\eta}_{r1} \left [ {\kappa_{\theta_{14}}\over \langle
\kappa_s(\gamma_1) \kappa_s(\gamma_4)\rangle}  \right ]
C^{\eta}_{s1} \left [ {\var \over \one}\right ]^{p+q+r+s-4}.
\end{eqnarray}
Finally the generating function  ${}_4\beta_{\kappa}^{\rm
star}(y_1,y_2,y_3,y_4)$ can be written as:
\begin{eqnarray}
 {}_4\beta_{\kappa}^{\rm star}(y_1,y_2,y_3,y_4) & = & { 1 \over \left
[\kappa_m \right ] ^2} \int_0^{\chi_s} d \chi \left [ {\corr \over
\langle \kappa_s(\gamma_1)\kappa_s(\gamma_2) \rangle } \right ]
\left [ {\kappa_{\theta_{13}} \over \langle
\kappa_s(\gamma_1)\kappa_s(\gamma_3) \rangle } \right ] \left [ {
\kappa_{\theta_{14}} \over \langle
\kappa_s(\gamma_1)\kappa_s(\gamma_4) \rangle} \right ] \left
[{\kappa_{\theta_0} \over \one} \right ]^{-4}
 \nonumber \\&&
\times \mu_3^{\eta} \left [ {y_1 } { \omega(\chi) \over r^2(\chi)
} {\var \over \one} \right ]\mu_1^{\eta} \left [ {y_2 } {
\omega(\chi) \over r^2(\chi) } {\var \over \one} \right ]
\mu_1^{\eta} \left [ {y_3 } { \omega(\chi) \over r^2(\chi) } {\var
\over \one} \right ]\mu_1^{\eta} \left [ {y_4 } { \omega(\chi)
\over r^2(\chi) } {\var \over \one} \right ]. \end{eqnarray}
 We can now change variables from $\kappa_s$ to $\eta_s$ and write:
\begin{eqnarray}
 {}_4\beta_{\eta}^{\rm star}(y_1,y_2,y_3,y_4) &  = & { 1 \over \left
[\kappa_m \right ] ^2} \int_0^{\chi_s} d \chi \left [ {\corr \over
\langle \kappa_s(\gamma_1)\kappa_s(\gamma_2) \rangle } \right ]
\left [ {\corri \over \langle \kappa_s(\gamma_2)\kappa_s(\gamma_3)
\rangle } \right ] \left [ { \corrj \over \langle
\kappa_s(\gamma_3)\kappa_s(\gamma_4) \rangle} \right ] \left
[{\kappa_{\theta_0} \over \one} \right ]^{-4} \nonumber \\&&
\times \mu_3^{\eta} \left [\kappa_m {y_1 } { \omega(\chi) \over
r^2(\chi) } {\var \over \one} \right ]\mu_1^{\eta} \left [\kappa_m
{y_2 } { \omega(\chi) \over r^2(\chi) } {\var \over \one} \right ]
\mu_1^{\eta} \left [\kappa_m {y_3 } { \omega(\chi) \over r^2(\chi)
} {\var \over \one} \right ]\mu_1^{\eta} \left [\kappa_m {y_4 } {
\omega(\chi) \over r^2(\chi) } {\var \over \one} \right ].
\end{eqnarray}
The four-point joint probability distribution function can not be
factorized as in the case of underlying mass distribution, but
using the same approximation as we have deployed throughout we can
simplify these terms and write
\begin{eqnarray}
 {}_4\beta_{\eta}^{\rm snake}(y_1,y_2,y_3,y_4) & = &
 \mu_1^{\eta}(y_1)\mu_2^{\eta}(y_2)\mu_2^{\eta}(y_3)\mu_1^{\eta}(y_4) \nonumber \\
 {}_4\beta_{\eta}^{\rm star}(y_1,y_2,y_3,y_4) & = &
 \mu_3^{\eta}(y_1)\mu_1^{\eta}(y_2)\mu_1^{\eta}(y_3)\mu_1^{\eta}(y_4),
\end{eqnarray}
which implies that, as in the lower-order reduced cumulant
correlators, we can write
\begin{equation}
C^{\eta}_{p111} = (\kappa_m)^{p+1} C^{\kappa}_{p111}.
\end{equation}
The joint probability distribution function for the convergence
field and the reduced convergence field now can be written
\begin{eqnarray}
p^{\kappa}(\kappa_1,\kappa_2,\kappa_3,\kappa_4)d\kappa_1 d\kappa_2
d\kappa_3 d\kappa_4 &  = &  p^{\kappa}(\kappa_1)
p^{\kappa}(\kappa_2) p^{\kappa}(\kappa_3) p^{\kappa}(\kappa_4) [ 1
+ b^{\kappa}(\kappa_1)\xi^{\kappa}_{12} b^{\kappa}(\kappa_2) +
{\rm cyc. perm.} \nonumber\\
 && + b^{\kappa}(\kappa_1) \xi^{\kappa}_{12} \nu^{\kappa}_2(\kappa_2)
\xi^{\kappa}_{23}\nu^{\kappa}_2(\kappa_3)
\xi^{\kappa}_{34}b(\kappa_4) + {\rm cyc. perm.}\nonumber\\ & & +
\nu^{\kappa}_1(\kappa_1) \xi^{\kappa}_{14} b^{\kappa}(\kappa_2)
\xi^{\kappa}_{24}b^{\kappa}(\kappa_3)
\xi^{\kappa}_{34}b^{\kappa}(\kappa_ 3)\nu^{\kappa}_3(\kappa_4) +
{\rm cyc. perm.} ] d\kappa_1 d\kappa_2 d\kappa_3 d\kappa_4
\nonumber \\ p^{\eta}(\eta_1,\eta_2,\eta_3,\kappa_4)d\eta_1
d\eta_2 d\eta_3 d\eta_4 & = & p^{\eta}(\eta_1) p^{\eta}(\eta_2)
p^{\eta}(\eta_3) p^{\eta}(\eta_4)[1 +
b^{\eta}(\eta_1)\xi^{\eta}_{12} b^{\eta}(\eta_2) + {\rm cyc.
perm.}  \nonumber \\&&+ b^{\eta}(\eta_1) \xi^{\eta}_{12}
\nu^{\eta}_2(\eta_2) \xi^{\eta}_{23}\nu^{\eta}_2(\eta_3)
\xi^{\eta}_{34}b^{\eta}(\eta_4) + {\rm cyc. perm.} \nonumber\\ &&
+ \nu_1^{\eta}(\eta_1) \xi_{14} b^{\eta}(\eta_2)
\xi^{\eta}_{24}b^{\eta}(\eta_3) \xi^{\eta}_{34}b^{\eta}(\eta_
3)\nu^{\eta}_3(\eta_4) + {\rm cyc. perm.}] d\eta_1 d\eta_2 d\eta_3
d\eta_4.
\end{eqnarray}

The generating function for reduced cumulant correlator $C_{p11}$,
i.e. $\mu_2(y)$ and $\nu_2^{\eta}(x)$ can be related by
\begin{equation}
\nu_3^{\eta}(x) h^{\eta}(x) = -{ 1 \over 2 \pi i}
\int_{-i\infty}^{i\infty} dy \mu_3^{\eta} (y) \exp (xy),
\end{equation}
where the scaling variable $x$ has the same meaning as defined
earlier. One can also derive a similar expression for the
cumulative $\nu_3^{\kappa}(>x)$, i.e.  $\nu_3^{\eta}(>x)$ beyond a
certain threshold; see Munshi et al. (1999a,b,c) for details.
Using the fact that
 $p(\kappa) = {p(\eta)/ \kappa_{m}}$;
$b(\kappa) = {b(\eta)/\kappa_{m}}$; $\nu_2(\kappa) =
{\nu_2(\eta)/\kappa_{m}}$  and $\xi_{\kappa} = {\xi_{\eta}/
\kappa_m^2}$ we can finally obtain
\begin{equation}
\nu_3^{\kappa}(\kappa) = {\nu_3^{\eta}(\eta) \over {\kappa_m}^3}.
\end{equation}
Since the scaling function $\nu_3(>x)$ encodes  information on the
kurtosis kurtosis of collapsed objects beyond certain threshold,
or hot-spots in convergence maps, it is interesting to notice how
such a quantity is directly related to the four-point cumulant
correlators (``star''-diagrams) of the background convergence
field. We can write the kurtosis $S_4(>\kappa_s)$ of those spots
in convergence maps which cross certain threshold $\kappa_s$ as:
\begin{equation}
S_4^{\kappa}(>\kappa_s) = 4{\nu_3^{\kappa}(> \kappa_s) \over b^{\kappa}(> \kappa_s)^3}+12{\nu_2^{\kappa}(>
 \kappa_s)^2 \over b^{\kappa}(> \kappa_s)^4}  = 4{\nu_3^{\eta}(>\eta_s)\over b^{\eta}(>
 \eta_s)^3} + 12{\nu_2^{\eta}(> \eta_s)^2 \over b^{\eta}(>
 \eta_s)^4}.
\end{equation}
As in the case of skewness, the kurtosis associated with points
which cross certain threshold in convergence map will have
precisely identical statistical properties as in the underlying
density field.

\section{Five-Point Statistics of Collpased objects from Convergence Maps}

The five-point cumulant correlators can be related to the fifth
order moments of collapsed objects in a very similar manner to the
preceding calculations, so we will simply  quote the results of
such an analysis in this section. The complication here is that,
in addition to having star and snake topologies, there is an
additional hybrid topology in this case.

For {\em star} topologies, the appropriate integrals will  involve
five node functions. Four of these will be of the form $\nu_1$,
and one $\nu_4$. There will also be four links between these nodes
denoting the correlations. Hence we can write
\begin{eqnarray}
 {}_5\beta_{\kappa}^{\rm star}(y_1,y_2,y_3,y_4,y_5) & = &
 \int_0^{\chi_s} r^2(\chi) d \chi \left [
{\kappa_{\theta_{15}} \over \langle
\kappa_s(\gamma_1)\kappa_s(\gamma_5) \rangle } \right ]
\left [ {\kappa_{\theta_{25}} \over \langle \kappa_s(\gamma_2)\kappa_s(\gamma_5)
\rangle }
\right ] \left [ { \kappa_{\theta_{35}} \over \langle
\kappa_s(\gamma_3)\kappa_s(\gamma_5) \rangle} \right ]\left [ { \kappa_{\theta_{45}} \over \langle
\kappa_s(\gamma_4)\kappa_s(\gamma_5) \rangle} \right ] \nonumber  \\&&
  \times \left [{\kappa_{\theta_0}\over  \one } \right ]^{-5}
\mu_4^{\eta} \left [ {y_1 } { \omega(\chi) \over r^2(\chi) } {\var
\over \one} \right ]\mu_1^{\eta} \left [ {y_2 } { \omega(\chi)
\over r^2(\chi) } {\var \over \one} \right ]\mu_1^{\eta} \left [ {y_3
 } { \omega(\chi) \over r^2(\chi) } {\var \over \one} \right
] \nonumber  \\&& \times \mu_1^{\eta} \left [ {y_4 } { \omega(\chi) \over r^2(\chi) } {\var \over \one} \right ]\mu_1^{\eta} \left [
] {y_5
] } { \omega(\chi) \over r^2(\chi) } {\var \over \one} \right ].
\end{eqnarray}

For a {\em snake} topology the relevant contributions will come
from three generating functions of three-point reduced cumulant
correlators $C_{p11}$ denoted as before by $\mu_2(y)$, and two
generating function of two-point reduced cumulant correlators
$C_{p1}$ which we have denoted by $\mu_1(y)$:
\begin{eqnarray}
 {}_5\beta_{\kappa}^{\rm snake}(y_1,y_2,y_3,y_4,y_5) & = &
\int_0^{\chi_s} r^2(\chi) d \chi \left [
{\kappa_{\theta_{12}} \over \langle
\kappa_s(\gamma_1)\kappa_s(\gamma_2) \rangle } \right ]
\left [ {\kappa_{\theta_{23}} \over \langle
\kappa_s(\gamma_2)\kappa_s(\gamma_3)\rangle } \right ]
\left [ { \kappa_{\theta_{34}} \over \langle
\kappa_s(\gamma_3)\kappa_s(\gamma_4) \rangle} \right ]
\left [ { \kappa_{\theta_{45}} \over \langle
\kappa_s(\gamma_4)\kappa_s(\gamma_5) \rangle} \right ] \nonumber  \\&&
\times \left [{\kappa_{\theta_0}\over  \one } \right ]^{-5}
\mu_1^{\eta} \left [ {y_1  } { \omega(\chi) \over r^2(\chi) } {\var
\over \one} \right ]\mu_2^{\eta} \left [ {y_2 } { \omega(\chi)
\over r^2(\chi) } {\var \over \one} \right ]  \mu_2^{\eta} \left [ {y_3 } { \omega(\chi) \over r^2(\chi) } {\var \over \one} \right ]
\nonumber \\ && \times \mu_2^{\eta} \left [ {y_4 } { \omega(\chi) \over r^2(\chi) } {\var \over \one} \right ]\mu_1^{\eta} \left [ {y_5
} { \omega(\chi) \over r^2(\chi) } {\var \over \one} \right ]. \end{eqnarray}

For the hybrid topology the contributions will come from the
generating function of four-point reduced cumulant correlators
$C_{p111}$ denoted as before by $\mu_2(y)$, three-point reduced
cumulant correlator $C_{p11}$ and the rest of the nodes $\mu_2(y)$
will be generating function for two-point cumulant correlators
$C_{p1}$. So finally we can write
\begin{eqnarray}
 {}_5\beta_{\kappa}^{\rm hybrid}(y_1,y_2,y_3,y_4,y_5) & = &
\int_0^{\chi_s} r^2(\chi) d \chi \left [ {\kappa_{\theta_{13}}
\over \langle \kappa_s(\gamma_1)\kappa_s(\gamma_3) \rangle }
\right ] \left [ {\kappa_{\theta_{23}} \over \langle
\kappa_s(\gamma_2)\kappa_s(\gamma_3) \rangle } \right ] \left [ {
\kappa_{\theta_{34}} \over \langle
\kappa_s(\gamma_3)\kappa_s(\gamma_4) \rangle} \right ]\left [ {
\kappa_{\theta_{45}} \over \langle
\kappa_s(\gamma_3)\kappa_s(\gamma_4) \rangle} \right ] \nonumber
\\&&\times \left [{\kappa_{\theta_0}\over  \one } \right ]^{-5}
\mu_1^{\eta} \left [ {y_1 } { \omega(\chi) \over r^2(\chi) } {\var
\over \one} \right ]\mu_1^{\eta} \left [ {y_2 } { \omega(\chi)
\over r^2(\chi) } {\var \over \one} \right ] \mu_3^{\eta} \left [
{y_3
 } { \omega(\chi) \over r^2(\chi) } {\var \over \one} \right
]\nonumber  \\&& \times \mu_2^{\eta} \left [ {y_4 } { \omega(\chi) \over r^2(\chi) }
{\var \over \one} \right ]\mu_1^{\eta} \left [ {y_5 } { \omega(\chi) \over r^2(\chi) }
{\var \over \one} \right ].
\end{eqnarray}

Making a change of variables again from the convergence field
$\kappa_s$ to the reduced convergence field $\eta_s$ we obtain the
following expressions for the five-point generating functions
which, again, cannot be factorised. For the star topology we have
\begin{eqnarray}
\beta_{\eta}^{\rm star}(y_1,y_2,y_3,y_4,y_5) & = & { 1 \over \left
[\kappa_m \right ] ^2} \int_0^{\chi_s} r^2(\chi) d \chi \left [
{\kappa_{\theta_{15}} \over \langle
\kappa_s(\gamma_1)\kappa_s(\gamma_5) \rangle } \right ] \left [
{\kappa_{\theta_{25}} \over \langle
\kappa_s(\gamma_2)\kappa_s(\gamma_5) \rangle } \right ] \left [ {
\kappa_{\theta_{35}} \over \langle
\kappa_s(\gamma_3)\kappa_s(\gamma_5) \rangle} \right ]\left [ {
\kappa_{\theta_{45}} \over \langle
\kappa_s(\gamma_4)\kappa_s(\gamma_5) \rangle} \right ] \nonumber
\\&& \times \left [{\kappa_{\theta_0}\over  \one } \right ]^{-5}
\mu_1^{\eta} \left [\kappa_m {y_1 } { \omega(\chi) \over r^2(\chi)
} {\var \over \one} \right ]\mu_1^{\eta} \left [ \kappa_m {y_2 } {
\omega(\chi) \over r^2(\chi) } {\var \over \one} \right ]
\mu_1^{\eta} \left [\kappa_m {y_3 } { \omega(\chi) \over r^2(\chi)
} {\var \over \one} \right ] \nonumber \\&& \times \mu_1^{\eta}
\left [ \kappa_m {y_4 } { \omega(\chi) \over r^2(\chi) } {\var
\over \one} \right ]\mu_4^{\eta} \left [ \kappa_m {y_5 } {
\omega(\chi) \over r^2(\chi) } {\var \over \one} \right ].
\end{eqnarray}
For the snake topology the relevant expression is
\begin{eqnarray}
\beta_{\eta}^{\rm snake}(y_1,y_2,y_3,y_4,y_5) & = & { 1 \over
\left [\kappa_m \right ] ^2} \int_0^{\chi_s} r^2(\chi) d \chi
\left [ {\corr \over \langle \kappa_s(\gamma_1)\kappa_s(\gamma_2)
\rangle } \right ] \left [ {\corri \over \langle
\kappa_s(\gamma_2)\kappa_s(\gamma_3) \rangle } \right ] \left [ {
\corrj \over \langle \kappa_s(\gamma_3)\kappa_s(\gamma_4) \rangle}
\right ]\left [ { \kappa_{\theta_{45}} \over \langle
\kappa_s(\gamma_4)\kappa_s(\gamma_5) \rangle} \right ] \nonumber
\\&& \times\left [{\kappa_{\theta_0}\over  \one } \right ]^{-5}
\mu_1^{\eta} \left [\kappa_m {y_1 } { \omega(\chi) \over r^2(\chi)
} {\var \over \one} \right ]\mu_2^{\eta} \left [\kappa_m {y_2 } {
\omega(\chi) \over r^2(\chi) } {\var \over \one} \right ]
\mu_2^{\eta} \left [\kappa_m {y_3 } { \omega(\chi) \over r^2(\chi)
} {\var \over \one} \right ] \nonumber \\&& \times \mu_2^{\eta}
\left [\kappa_m {y_4 } { \omega(\chi) \over r^2(\chi) } {\var
\over \one} \right ] \mu_1^{\eta} \left [ \kappa_m {y_5 } {
\omega(\chi) \over r^2(\chi) } {\var \over \one} \right ],
\end{eqnarray}
and similarly for the hybrid topology:
\begin{eqnarray}
\beta_{\eta}^{\rm hybrid}(y_1,y_2,y_3,y_4,y_5) & = & { 1 \over
\left [\kappa_m \right ] ^2} \int_0^{\chi_s} r^2(\chi) d \chi
\left [ {\kappa_{\theta_{13}} \over \langle
\kappa_s(\gamma_1)\kappa_s(\gamma_3) \rangle } \right ] \left [
{\kappa_{\theta_{23}} \over \langle
\kappa_s(\gamma_2)\kappa_s(\gamma_3) \rangle } \right ] \left [ {
\kappa_{\theta_{34}} \over \langle
\kappa_s(\gamma_3)\kappa_s(\gamma_4) \rangle} \right ]\left [ {
\kappa_{\theta_{45}} \over \langle
\kappa_s(\gamma_4)\kappa_s(\gamma_5) \rangle} \right ] \nonumber
\\&& \times \left [{\kappa_{\theta_0}\over  \one } \right ]^{-5}
\mu_1^{\eta} \left [ \kappa_m {y_1 } { \omega(\chi) \over
r^2(\chi) } {\var \over \one} \right ]\mu_1^{\eta} \left [
\kappa_m {y_2 } { \omega(\chi) \over r^2(\chi) } {\var \over \one}
\right ]  \mu_3^{\eta} \left [ \kappa_m {y_3 } { \omega(\chi)
\over r^2(\chi) } {\var \over \one} \right ] \nonumber \\&& \times
\mu_2^{\eta} \left [ \kappa_m {y_4 } { \omega(\chi) \over
r^2(\chi) } {\var \over \one} \right ]\mu_1^{\eta} \left [
\kappa_m {y_5 } { \omega(\chi) \over r^2(\chi) } {\var \over \one}
\right ]. \end{eqnarray}

Using the same approximations as we have used before for the
lower-order cumulants, we can see that the generating function
$\nu_4(y)$ and its associated scaling function $\nu_4(x)$ for the
reduced convergence field are, once again,  exactly the same as
the one for underlying mass distribution. This leads to
\begin{eqnarray}
{}_5\beta_{\eta}^{\rm star}(y_1,y_2,y_3,y_4,y_5) & = &
\mu_1^{\eta}(y_1) \mu_1^{\eta}(y_2) \mu_1^{\eta}(y_3)
\mu_4^{\eta}(y_4) \mu_1^{\eta}(y_5), \nonumber \\
{}_5\beta_{\eta}^{\rm snake}(y_1,y_2,y_3,y_4,y_5) & = &
\mu_1^{\eta}(y_1) \mu_2^{\eta}(y_2) \mu_2^{\eta}(y_3)
\mu_2^{\eta}(y_4) \mu_1^{\eta}(y_5),\nonumber \\
{}_5\beta_{\eta}^{\rm hybrid}(y_1,y_2,y_3,y_4,y_5) & = &
\mu_1^{\eta}(y_1) \mu_1^{\eta}(y_2) \mu_3^{\eta}(y_3)
\mu_2^{\eta}(y_4) \mu_1^{\eta}(y_5).
\end{eqnarray}
As in the case of the fourth-order cumulant correlators, these
results now can be used to express the five-point joint
probability distribution function for both convergence and reduced
convergence field. Clearly the new vertex $\mu_4(y)$ and
associated scaling function $\nu_4(x)$ (or the analogous
$\nu_4(\eta_s)$ or $\nu_4(\kappa_s)$) can be used to express the
fifth-order cumulant for collapsed objects:
\begin{equation}
\nu_4^{\kappa}(\kappa) = {\nu_4^{\eta}(\eta) \over \kappa_m^4}.
\end{equation}

Finally the fifth order one-point moment of the collapsed objects
is
\begin{equation}
S_5^{\kappa}(>\kappa_s) = 5{\nu_4^{\kappa}(> \kappa_s) \over b^{\kappa}(> \kappa_s)^4}+60{\nu_3^{\kappa}(>
 \kappa_s) \nu_2^{\kappa}(\kappa_s)^2 \over b^{\kappa}(> \kappa_s)^4}+
 60{{\nu_2^{\kappa}}(>\kappa_s)^3
 \over b(>\kappa_s)^4}  =  5{\nu_4^{\eta}(> \eta_s) \over b^{\eta}(> \eta_s)^4}+60^{\eta}{\nu_3^{\eta}(>\eta_s) \nu_2^{\eta}(>\eta_s)
 \over b^{\eta}(> \eta_s)^4}+ 60{\nu_2^{\eta}(>\eta_s)^3 \over b^{\eta}(>\kappa_s)^4}.
\end{equation}

\begin{figure}
\protect\centerline{
 \epsfysize = 1.8truein
 \epsfbox[270 6 377 192]
 {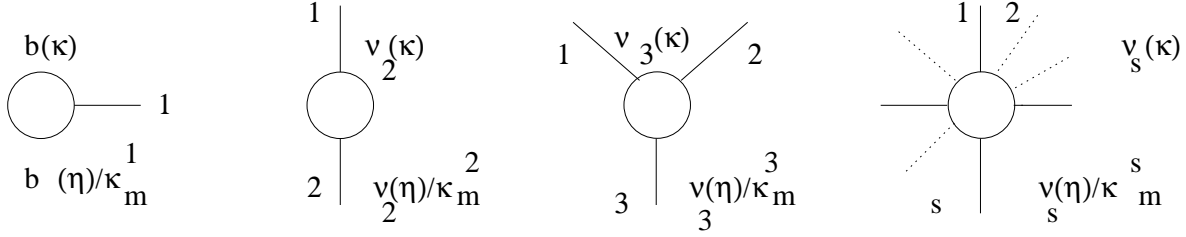} }
 \caption{Transformation of nodes representing bright spots in
 convergence maps for a change of variables from convergence
 $\kappa_s$ to reduced convergence $\eta_s$. However such a
 decomposition is possible only under certain simplifying assumption
 (see text for details).}
\end{figure}

\section{Generalization to Arbitrary Order Statistics}

Let us now put together what we have learned so far and explain
what can be learned about correlations of arbitrary order. If we
consider a diagram of arbitrary topology and of arbitrary order it
will consist of different kinds of nodes. These nodes will be
represented by associated  scaling functions $\mu_s(y)$ and act
as generating functions for the reduced cumulant correlators of
various order $C_{p\dots1}$. The term will also consist of
products of the factors which represent correlation function
associated with the links which join these nodes. Products of
these terms when integrated along the line of sight direction will
provide us the joint generating function of arbitary order for
convergence field $\kappa_s$:
\begin{equation}
{}_{N}\beta_{\kappa}^{\rm topology}(y_1, \dots, y_s) = \int
{r^2(\chi) d \chi }\left [ {\var \over \langle \kappa_s^2(\gamma)
\rangle } \right ]^{- N} \prod_{\rm all~links~(i,j)}^{N-1} \left [
{\kappa_{\theta}}_{ij} \over \langle \kappa(\gamma_i)
\kappa(\gamma_j)\rangle \right ]  \prod_{\rm all~nodes~(k)}^{N}
\mu_s^{\eta} \left [ y_k {\omega(\chi)\over r^2(\chi)} {\var \over
\langle \kappa_s^2(\gamma) \rangle } \right ].
\end{equation}
Changing the variables from convergence field $\kappa_s$ to
reduced convergence $\eta_s$ we can  write
\begin{equation}
{}_{N}\beta_{\eta}^{\rm topology}(y_1, \dots, y_s) = { 1 \over
\left [\kappa_m \right ] ^2} \int { r^2(\chi) d \chi}\left [ {\var
\over \langle \kappa_s^2(\gamma) \rangle } \right ]^{- N}
\prod_{\rm all~links~(i,j)}^{N-1} \left [ {\kappa_{\theta}}_{ij}
\over \langle \kappa(\gamma_i) \kappa(\gamma_j)\rangle \right ]
\prod_{\rm all~nodes~(k)}^{N} \mu_s^{\eta} \left [ y_k \kappa_m
{\omega(\chi)\over r^2(\chi)} {\var \over \langle
\kappa_s^2(\gamma) \rangle } \right ].
\end{equation}
As before, this  is the most general expression and includes all
other results which we have obtained so far. We want to emphasize
 that deriving this result we have not used any
approximation other than that of small angles  and these integrals
can be computed numerically to obtain the joint multi-point PDF
for convergence fields.

Given a specific model for the underlying mass distribution we can
now  relate all the multi-point properties of the projected
density field  to be obtained from weak lensing surveys. However,
as we indicated before,  with some approximations it is possible
to simplify these results still further. This gives us an
interesting insight into the statistical nature of the underlying
mass distribution:
\begin{equation}
{}_{N}\beta_{\eta}^{\rm topology}(y_1, \dots, y_s) = \prod_{\rm
all~nodes~(k)}^{N} \mu_s^{\eta} [ y_k]
\end{equation}
These approximations have already been tested against the results
of high resolution  ray tracing experiments (Munshi \& Jain
1999a,b; Munshi 1999)which have confirmed that these
simplifications indeed reproduce the results in a remarkably
accurate way.

The vertices associated with collapsed objects $\mu_n(y)$, can be
linked to generating functions for the vertices in the matter
correlation hierarchy in a very interesting way. This has been
done in an order-by-order expansion (Bernardeau \& Schaeffer 1992;
Munshi, Coles \& Melott 1999a) and also to arbitrary order in a
non-perturbative approach (Bernardeau \& Schaeffer 1999). These
techniques were developed to compute the joint multi-point
statistics of collisionless clustering. Our study shows that the
statistics of reduced convergence field are deeply linked to the
statistics of underlying mass distribution. However, we should
keep in mind that although the same functional form for $\nu_n(x)$
holds for both reduced convergence field and the underlying mass
distribution the variance associated with them are very different.
The variance for the mass distribution is very high on small
length scales, while the variance of reduced convergence field is
small. This  means that while it is possible to compute asymptotic
forms for the scaling functions (such as  $\nu_n(x)$) for very
small variance,  complete information about $\nu_n(x)$ can only be
obtained by a numerical integrations. This has already been done
by Munshi \& Jain (1999a), Munshi (2000) for the case of PDF and bias.

\section{Discussion}

In this paper we have explored the statistics of regions where the
lensing convergence exceeds some threshold value. Our main purpose
in this is to demonstrate the deep connection between statistical
descriptors of the convergence and those of the  underlying mass.

Ongoing weak lensing surveys with wide field CCD are likely to
produce shear maps on areas of order 10 square degrees or less;
existing feasibility studies cover relatively small angles (Bacon
et al. 2000; van Waerbeke et al. 2000). In the future, areas
larger than $10^{\circ} \times 10^{\circ}$ are also feasible, e.g.
from the MEGACAM camera on the Canada France Hawaii Telescope and
the VLT Survey Telescope. Ongoing optical (SDSS; Stebbins et al.
1997) and radio (FIRST; Refregier et al. 1997) surveys can also
provide useful imaging data for weak lensing surveys. Such surveys
will provide maps of the projected density of the universe and
thus will help us both to test different dark matter models and
probe the background geometry of the universe. They will also
provide a unique opportunity to test different {\em ansatze} for
gravitational clustering in the highly non-linear regime in an
unbiased way. Traditionally such studies have used galaxy surveys,
with the disadvantage that galaxies are biased tracers of the
underlying mass distribution.

Most previous studies in weak lensing statistics used a
perturbative formalism, which is applicable in the quasilinear
regime and thus requires large smoothing angles. To reach the
quasi-linear regime, survey regions must exceed areas of order 10
square degrees. Since existing CCD cameras typically have
diameters of $0.25^{\circ}-0.5^{\circ}$, the initial weak lensing
surveys are likely to provide us statistical information on small
smoothing angles, of order $10'$ and less. This makes the use of
perturbative techniques a serious limitation, as the relevant
physical length scales  are in the highly nonlinear regime. In our
earlier studies (Munshi \& Coles 2000a) we have shown how  lower
order moments, such as cumulants and cumulant correlators, can
also be extended to the case of weak lensing surveys. Although an
important step towards understanding the physics of gravitational
clustering and constraining the cosmological parameters, this work
did not represent a complete picture of the entire PDF and
multi-point generalisations of the cumulants and cumulant
correlators. This was another aim of the present study. Extending
our earlier results we have shown that indeed such results can be
obtained once we assume a hierarchical model for the underlying
mass distribution.

Numerical comparison  of the one point probability distribution
function (PDF) and bias of the convergence field have already been
carried out by Munshi \& Jain (1999a,b). The success of analytical
results presented here indicates that the PDF, bias and other
higher order statistics for ``hot-spots'' in convergence maps for
a desired cosmological model can be computed as a function of
smoothing angle and redshift distribution. Thus physical effects,
such as the magnification distribution for Type Ia supernovae, can
be conveniently computed. Thus we have a complete analytical
description, based on models for gravitational clustering, for the
full set of statistics of interest for weak lensing -- the one
point pdf, two point correlations, and the hierarchy of higher
order cumulants and cumulant correlators. This analytical
description has powerful applications in making predictions for a
variety of models and varying the smoothing angle and redshift
distribution of source galaxies; numerical studies are far more
limited in the parameter space that can be explored.

The lower-order statistics of
 ``hot-spots'' in the convergence field have also already been studied by (Munshi \& Jain 1999a,b).
 They demonstrated that, given a threshold, it is  possible to
link the statistics of collapsed objects with the `hot-spots' in
convergence field. It was also established that both one-point
objects (such as PDF) and two-point objects (such as bias)
measured from ray tracing experiments reproduce the analytical
predictions with very good accuracy. This strongly motivates this
present study, which develops these ideas further. Clearly the
hierarchical {\em ansatz} has not been tested against N-body
simulation in its full generality. While we have a very good
analytical models for one-point scaling functions (Munshi et al.
1999a,b) which have been tested against numerical simulations,
corresponding studies for two-point quantities  are still lacking.
One expects that situation will improve with availability of
larger N-body simulations.

As we found that a suitable transformations of variables from
convergence to reduced convergence can make the analysis very
simple. Moreover we found that the reduced convergence has exactly
the same statistics as the underlying mass distribution under some
simplifying assumptions. However, in general the generating
functions do not obey a factorization property similar to their
counterpart for underlying mass distribution due to a line of
sight integration. In general the minimum value of $\kappa$ i.e.
$\kappa_m$ plays an important role in the transformation.

The dependence of convergence statistics on parameters of the
cosmological background model enters through their effect on
$\kappa_m$. This is why  the statistics of $\eta_s$ are not
sensitive to the background geometry or dynamics, but the
statistics of convergence $\kappa$ are very sensitive to
parameters such as $\Omega$ and $\Lambda$. This implies that it is
possible to determine or constrain these parameters using weak
lensing surveys. Numerical evaluation of this approach has already
been done by Munshi \& Jain (1999a,b) and Munshi (2000)
who found very good agreement between analytical predictions and
numerical simulations. While the statistics of the  convergence
field is very sensitive to the cosmological parameters, the $S_N$
parameters  of bright spots which correspond to the collapsed
objects in mass distribution are independent of such parameters
and are approximately the same as those of collapsed objects; the
bias of these bright-spots does depend on cosmplogical parameters.

Our studies also indicate that several approximations involved in
weak lensing studies are valid even in the highly non-linear
regime of observational interest. While a weakly clustered dark
matter distribution is expected to produce small deflections of
photon trajectories, it is not clear if the effect of highly
over-dense regions capable of producing large deflections can also
be modeled using a hierarchical {\em ansatz} and the weak lensing
approximation. Numerical studies have shown
 that this is indeed
the case; the effect of density inhomogeneities on very small
angular scales can be predicted with high accuracy using
analytical approximations, which confirms for example that the
Born approximation is valid in the highly nonlinear regime.

The finite size of weak lensing catalogs will play an important
role in the determination of cosmological parameters. It is
therefore of interest to incorporate such effects in future
analysis. Noise due to the intrinsic ellipticities of lensed
galaxies will also need to be modeled. The present study is based
on a top-hat window function which is easier to incorporate in
analytical computations. It is possible to extend our study to
other statistical estimators such as $M_{\rm ap}$, which use a
compensated filter to smooth the shear field and may be more
suitable for observational studies (Schneider et al 1997;
Reblinsky et al 1999). Our method of computing cumulants and
cumulant correlators or PDF and bias can also be generalized to
the case of other window functions and we hope to present such
results elsewhere.

It is interesting to consider an alternative to the hierarchical
ansatz for the distribution of dark matter in the highly nonlinear
regime. The dark matter can be modelled as belonging to haloes
with a mass function given by the Press-Schechter formalism, and
spatial distribution modeled as in Mo \& White (1996). When
supplemented by a radial profile for the dark halos such a
prescription can be used to compute the one-point cumulants of the
convergence field. Reversing the argument, given the one point
cumulants of the convergence field it is possible to estimate the
statistics of dark haloes.

\section*{Acknowledgment}
DM was supported by a Humboldt Fellowship at the Max Planck
Institut fur Astrophysik when this work was performed. It is a
pleasure for DM to acknowledge many helpful discussions with
Francis Bernardeau, Patric Valageas and Katrin Reblinsky. Some of the
analytical results which we have presented here were tested successfully
against ray tracing simulations in collaboration with Bhuvnesh Jain. It
is a pleasure for DM to thank him for several enjoyable collaborations.

\end{document}